\documentclass[prd,aps]{revtex4}

\begin{document}

\title{Normal forces in stationary spacetimes}

\author{Richard H.~Price} 
\affiliation{Department of Physics,
University of Utah, Salt Lake City, Utah 84112}

\begin{abstract}
\begin{center}
{\bf Abstract}
\end{center}
For geodesic motion of a particle in a stationary spacetime the $U_0$
component of particle 4-velocity is constant and is considered to be a
conserved mechanical energy. We show that this concept of a conserved
mechanical energy can be extended to particles that move under the
influence of a ``normal force,'' a force that, in the stationary
frame, is orthogonal to the motion of the particle. We illustrate
the potential usefulness of the concept with a simple example.
\end{abstract}

\maketitle

For a particle with 4-velocity
$\vec{U}$, moving on a geodesic in a 
spacetime with 
timelike Killing vector $\vec{\xi}$, it is well known that the quantity
\begin{equation}\label{defE} 
E\equiv \vec{U}\cdot{\vec\xi} 
\end{equation}
is conserved along the geodesic particle world line\cite{Wald84}. The
quantity $E$ can then be taken to be a conserved mechanical energy
(per unit particle mass), analogous to the classical mechanical energy
that includes gravitational potential energy.  In Newtonian mechanics
this concept can be extended to particles that move under the influence
of normal forces, forces that serve only to constrain the particle
motion to a geometric path. Newtonian normal forces  are orthogonal to the
velocity of the particle, and therefore perform no work on the
particle and do not change the mechanical energy. Here we show that
this can also be made to apply for relativistic gravitation.  We
extend the idea of a normal forces to stationary spacetime, and we
show that under the influence of such normal forces, the quantity $E$
of Eq.~(\ref{defE}) is conserved. Though this concept is quite simple,
basic, and potentially useful, it does not seem to appear in textbooks.

Our definition of normal force uses the operator $P_{\vec\xi}$ which 
projects any vector $\vec{w}$ to 
\begin{equation}\label{Pdef} 
P_{\vec{\xi}}(\vec{w})=\vec{w}-\left[\frac{\vec{\xi}\cdot\vec{w}
}{\vec{\xi}\cdot\vec{\xi}
}\right]\;\vec{\xi}\,,
\end{equation}
so that $P_{\vec{\xi}}(\vec{w})$ is orthogonal to $\vec\xi$.  Using
this operator, we take $P_{\vec{\xi}}(\vec{U})$ and
$P_{\vec{\xi}}(\vec{a})$ respectively to be the spatial velocity and
acceleration of the particle. Our condition that only normal forces
are acting on a particle is the condition that the particle's spatial
acceleration is orthogonal to its  spatial velocity, or
\begin{equation}\label{normalforcedef} 
P_{\vec{\xi}}(\vec{U})\cdot P_{\vec{\xi}}(\vec{a})=0
\ \ \mbox{normal force definition}\,.
\end{equation}
A straightforward substitution of Eq.~(\ref{Pdef}) into (\ref{normalforcedef})
gives
\begin{equation} 
P_{\vec{\xi}}(\vec{U})\cdot P_{\vec{\xi}}(\vec{a})=
\vec{U}\cdot\vec{a}-
\frac{(\vec\xi\cdot\vec{U})(\vec\xi\cdot\vec{a})
}{(\vec\xi\cdot\vec{\xi})}\,.
\end{equation}
Since $\vec{U}\cdot\vec{a}$ must be zero, and 
$\vec{\xi}\cdot\vec{U}$ cannot be zero, it follows that 
\begin{equation}
\vec\xi\cdot\vec{a}=0\,.
\end{equation}
We now take the derivative  $dE/d\tau$ where $E$ is the energy
defined in Eq.~(\ref{defE}) and
$\tau$ is proper time along the worldline:
\begin{equation}
\frac{dE}{d\tau}=\nabla_{\vec{U}}
\left(\vec{U}\cdot\vec\xi\right)=
\vec{a}\cdot{\vec\xi}+\vec{U}\cdot\nabla_{\vec{U}}\xi
\,.
\end{equation}
The last term $U^\alpha\xi_{\alpha;\beta}U^\beta$ vanishes if $\vec{\xi}$
is a Killing vector, and the first term vanishes if the particle 
is acted on only by normal forces. The energy $E$ is therefore constant
along the worldline.

This extended concept of conserved mechanical energy is most usefully
applied 
in a coordinate system with
$\vec\xi=\partial/\partial x^0$, and the metric coefficients
independent of time $x^0$.  (We use units in which $c=G=1$ and other
conventions of Misner {\em et al.}\cite{MTW}.)  We suppose that the
path to which the normal forces constrain the particle is specified
through $x^j(\lambda)$, where $x^j$ is a spatial coordinate and
$\lambda$ is some curve parameter. We suppose also that the energy
$E\equiv\vec{U}\cdot\vec{\xi}=U_0$ is specified. From
$\vec{U}\cdot\vec{U}=-1$ we have
\begin{equation}\label{UdotU} 
-1=
g_{00}\left(U^0\right)^2+g_{ij}\;\frac{dx^i}{d\tau}\;\frac{dx^j}{d\tau}\,.
\end{equation}
Here and below, summation over spatial indices $i,j,\ldots$ is understood.
Equation (\ref{UdotU}) gives us
\begin{equation}\label{UdotU2} 
-1=g_{00}^{-1}
\left(E-g_{0j}
\frac{dx^j}{d\lambda}\;\frac{d\lambda}{d\tau}\right)^2
+g_{ij}\;\frac{dx^i}{d\lambda}\;\frac{dx^j}{d\lambda}
\;\left(\frac{d\lambda}{d\tau}\right)^2
\,.
\end{equation}
In this equation, the metric coefficients $g_{00}$,$g_{0j}$, and
$g_{ij}$ are known functions of $x^j$, and hence of $\lambda$. The
only unknown in Eq.~(\ref{UdotU2}), then, is $d\lambda/d\tau$, and 
Eq.~(\ref{UdotU2}) is an ordinary  differential equation for $\lambda(\tau)$.
Once this is solved, $x^j(\tau)$, and hence the dynamics, is completely
determined. 

As a simple explicit example, for the Schwarzschild spacetime expressed
in the usual coordinates,
\begin{equation}
ds^2=-\left(1-2M/r\right)dt^2
+\left(1-2M/r\right)^{-1}
dr^2+r^2\left(d\theta^2+\sin^2\theta\,d\phi^2\right)
\end{equation}
we consider the  ``straight'' curve
$r\cos\phi=b$ 
in the $\theta=\pi/2$ equatorial plane. From $\vec{U}\cdot\vec{U}=-1$
we have 
\begin{equation}\label{eom} 
-1=-\,\frac{E^2}{1-2M/r}+\left[\frac{r^2
-b^2}{1-2M/r}\;\frac{r^2}{b^2}
+r^2\right]\;\left(\frac{d\phi}{d\tau}\right)^2\,.
\end{equation}
From this and $r=b/\cos{\phi}$, we can find $\phi(\tau)$, $r(\tau)$, $\phi(t)$, 
and so forth. 

It is particularly interesting to see the effect of the normal force
on $L\equiv U_\phi=r^2\,d\phi/d\tau$. For geodesic motion in an
azimuthally symmetric spacetime, $L$ is a conserved quantity
considered to be particle angular momentum per unit mass. For 
the nongeodesic motion along $r\cos\phi=b$, we have, from
Eq.~(\ref{eom}), that 
\begin{equation}\label{Leq} 
L^2=b^2\;\frac{E^2-1+2M/r
}{1-2Mb^2/r^3}\,.
\end{equation}
Since $L$ is not constant, we can conclude that the normal force
applies a torque to the particle.

\end{document}